\documentclass[%
reprint,
superscriptaddress,
 bibnotes,
 amsmath,amssymb,
 aps,
prl
]{revtex4-1}

\usepackage{graphicx}
\usepackage{dcolumn}
\usepackage{bm}
\usepackage{color}

\begin{document}


\title{Distinct fingerprints of charge density waves and electronic standing waves in ZrTe$_3$}

\author{Li~Yue}
\thanks{These authors contributed equally to this study.}
\affiliation{International Center for Quantum Materials, School of Physics, Peking University, Beijing 100871, China}
\author{Shangjie~Xue}
\thanks{These authors contributed equally to this study.}
\affiliation{Department of Physics, Massachusetts Institute of Technology, Cambridge, MA 02139, USA}
\author{Jiarui~Li}
\thanks{These authors contributed equally to this study.}
\affiliation{Department of Physics, Massachusetts Institute of Technology, Cambridge, MA 02139, USA}
\author{Wen~Hu}
\affiliation{National Synchrotron Light Source II, Brookhaven National Laboratory, Upton, New York 11973, USA}
\author{Andi~Barbour}
\affiliation{National Synchrotron Light Source II, Brookhaven National Laboratory, Upton, New York 11973, USA}
\author{Feipeng~Zheng}
\affiliation{Siyuan Laboratory, Guangzhou Key Laboratory of Vacuum Coating Technologies and New Energy Materials, Department of Physics, Jinan University, Guangzhou 510632, China}
\author{Lichen~Wang}
\affiliation{International Center for Quantum Materials, School of Physics, Peking University, Beijing 100871, China}
\author{Ji~Feng}
\affiliation{International Center for Quantum Materials, School of Physics, Peking University, Beijing 100871, China}
\affiliation{Collaborative Innovation Center of Quantum Matter, Beijing 100871, China}
\author{Stuart~B.~Wilkins}
\affiliation{National Synchrotron Light Source II, Brookhaven National Laboratory, Upton, New York 11973, USA}
\author{Claudio~Mazzoli}
\affiliation{National Synchrotron Light Source II, Brookhaven National Laboratory, Upton, New York 11973, USA}
\author{Riccardo~Comin}
\email[]{rcomin@mit.edu}
\affiliation{Department of Physics, Massachusetts Institute of Technology, Cambridge, MA 02139, USA}
\author{Yuan~Li}
\email[]{yuan.li@pku.edu.cn}
\affiliation{International Center for Quantum Materials, School of Physics, Peking University, Beijing 100871, China}
\affiliation{Collaborative Innovation Center of Quantum Matter, Beijing 100871, China}

\begin{abstract}
Experimental signatures of charge density waves (CDW) in high-temperature superconductors have evoked much recent interest, yet an alternative interpretation has been theoretically raised based on electronic standing waves resulting from quasiparticles scattering off impurities or defects, also known as Friedel oscillations (FO). Indeed the two phenomena are similar and related, posing a challenge to their experimental differentiation. Here we report a resonant X-ray diffraction study of ZrTe$_3$, a model CDW material. Near the CDW transition, we observe two independent diffraction signatures that arise concomitantly, only to become clearly separated in momentum while developing very different correlation lengths in the well-ordered state. Anomalously slow dynamics of mesoscopic ordered nanoregions are further found near the transition temperature, in spite of the expected strong thermal fluctuations. These observations reveal that a spatially-modulated CDW phase emerges out of a uniform electronic fluid via a process that is promoted by self-amplifying FO, and identify a viable experimental route to distinguish CDW and FO.

\end{abstract}

\maketitle

Charge density waves (CDW) and related phenomena have been a long-standing topic in condensed matter research \cite{GrunerRevModPhys1988,MonceauAdvPhys2012,RossnagelJPhysCondMat2011}. A renewed interest in this topic was brought about in recent years by the discovery of ubiquitous signatures of CDW in cuprate high-temperature superconductors \cite{KeimerNature2015,CominAnnualReview2016}. On the one hand, the formation of CDW in at least some of the cuprates is accompanied by sharp phonon-dispersion (Kohn) anomalies \cite{LeTaconNatPhys2014}, which are commonly found in conventional CDW systems. On the other hand, even though long-range CDW can be stabilized in the cuprates by a variety of external fields \cite{GreberScience2015,BluschkeNatcom2018,KimScience2019},
the three-dimensional ordering propagation vector is at odds with that of the Kohn anomalies in zero-field condition \cite{LeTaconNatPhys2014,KimScience2019}. In contrast,
the zero-field charge correlations are often found to be short-ranged \cite{TabisNatCom2014,CominScience2014,CominScience2015,CampiNature2015} and coexisting with a rather inhomogeneous electronic background \cite{McElroyNature2003,MesarosScience2011,KohsakaScience2007,CampiNature2015,KangNatPhys2019}. It is, therefore, of primary interest to elucidate the role of disorder during the incipience of the CDW state. In fact, it has been proposed that band structure effects \cite{AbbamontePhysicaC2012,TorreNatPhys2016,TorrePRB2016,TorreNewJPhys2015}, namely Friedel oscillations (FO) seeded by impurities and quenched disorder, could produce experimental signatures that look similar to genuine CDW \cite{TorreNatPhys2016,TorrePRB2016,
KangNatPhys2019}, so it is important to establish an experimental methodology to distinguish such contributions.

A main challenge in experimentally addressing the role of disorder pertains to the detection length scale. Disorder is known to provide a pinning potential to foster the stabilization of dynamical charge correlations near the transition point \cite{LeePRB1979,GrunerRevModPhys1988,TuckerPRB1988,MonceauAdvPhys2012,PinsollePRL2012,LebollochPRB2005}.
The presence of disorder may affect the correlation length and dynamics, and such effects are expected to occur at the mesoscopic scale, \textit{i.e.}, comparable to the CDW domain sizes which are often much greater than the density-modulation periodicity. Unfortunately, the majority of experiments carried out to date for addressing disorder effects in CDW materials fall either into the macroscopic regime, such as transport and thermodynamic measurements \cite{BiljakovicPRL1989,KrizaPRL1986,SinchenkoPRB2016}, or into the atomic-scale regime, such as scanning probe experiments \cite{NovelloPRB2015,IshiokaPRB2011}.

\begin{figure*}
\includegraphics[width=3.5in]{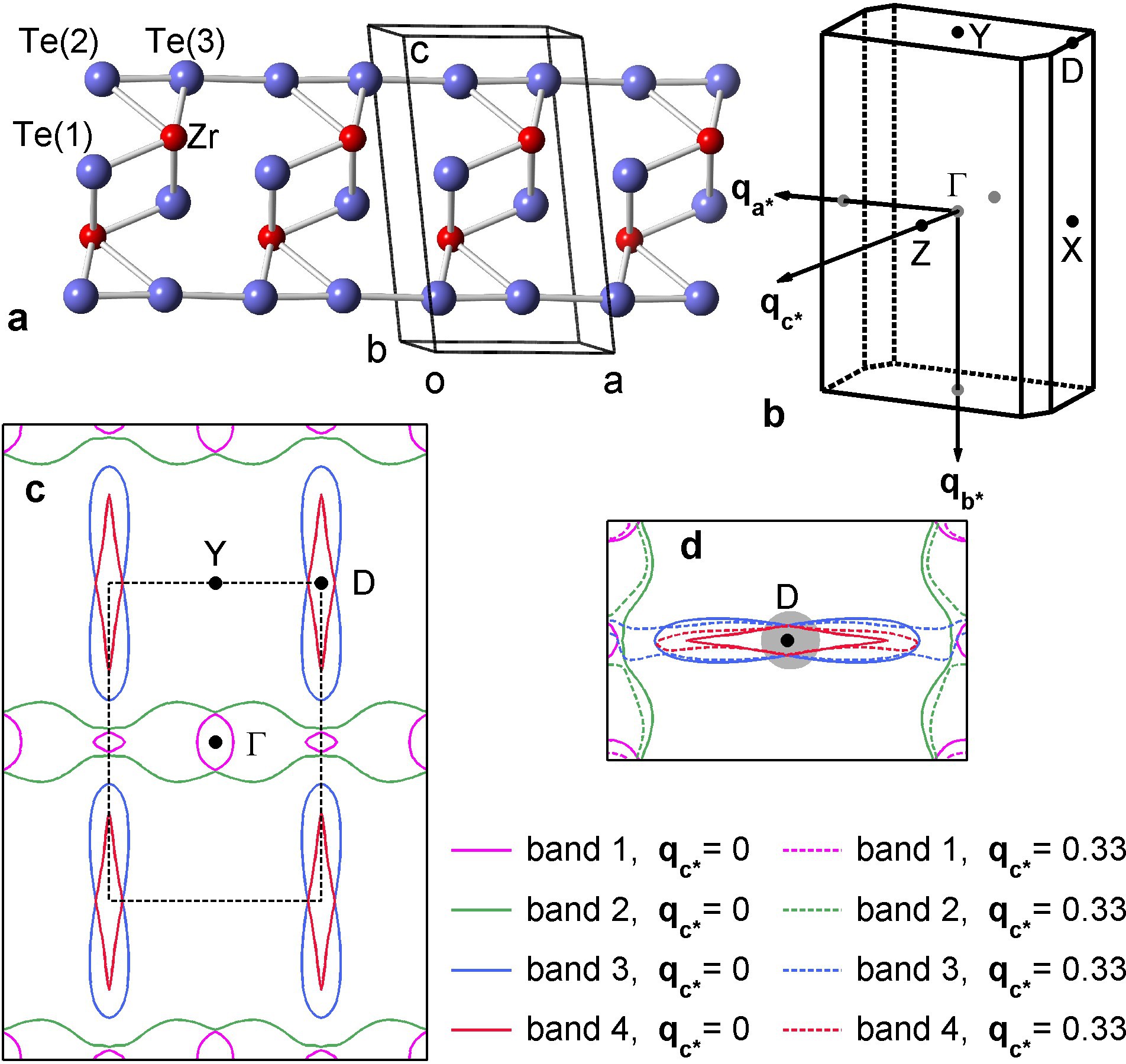}
\caption{\label{Fig1}
(a) Crystal structure of ZrTe$_3$. (b) The first Brillouin zone. (c) Calculated Fermi surface \cite{SM} at $\mathbf{q}_{\mathbf{c}^*}=0$. The Fermi surfaces indicated in red and blue are flat sheets that run along the $\mathbf{c}^*$ direction perpendicular to the plane of display. (d) Illustration of Fermi surface nesting properties. Solid and dashed lines refer to the Fermi surfaces at $\mathbf{q}_{\mathbf{c}^*}=0$ and $\mathbf{q}_{\mathbf{c}^*}=0.33$, respectively, where the difference indicates Fermi surface warping along $\mathbf{c}^*$. The electronic gap associated with the CDW order opens from the shaded region around the D-point.
}
\end{figure*}

For scattering experiments to survey the mesoscopic scale, special techniques are usually required. Here we use resonant soft X-ray diffraction to study a prototypical CDW material, ZrTe$_3$, with very high accuracy. The length-scale challenge is met on two fronts: first, we use soft X-rays in conjunction with a high-resolution area detector to achieve high momentum resolution, enabling us to distinguish Fourier signatures that are only slightly different in spatial periodicity. Second, we use a highly coherent X-ray beam to detect the domain texture and dynamics via interference patterns known as ``speckles'' \cite{
ShpyrkoNature2007,ChenPRL2016,PinsollePRL2012}. The experiment further benefits from resonant enhancement of diffraction signals at the Te $M$ absorption edges -- providing increased sensitivity to the very weak charge correlations near the CDW melting point even for relatively short acquisition times.

\section{Results}

\textbf{System.} ZrTe$_3$ is a quasi-one-dimensional metal belonging to the monoclinic space group $P2_1/m$ [see Fig.~\ref{Fig1}(a)], with $a=5.89$ \r{A}, $b=3.93$ \r{A}, $c=10.09$ \r{A}, $\alpha=\gamma=90^{\circ}$, $\beta=97.8^{\circ}$ \cite{
HuPRB2015,StoweJSolidStateChem1998}. According to resistivity measurements (Fig.~S1 in \cite{SM}), the CDW order develops below $T_\mathrm{CDW}\approx63$ K, with wave vector $\mathbf{q}_\mathrm{CDW} = (0.07 \mathbf{a}^*, 0, 0.33 \mathbf{c}^*)$. The Fermi surface, contributed by four bands, comprises two sectors [Fig.~\ref{Fig1}(b-d)]: three-dimensional (3D) pockets around the Brillouin zone (BZ) center $\Gamma$, and quasi-1D sheets running along the BZ boundary with Fermi velocity primarily in the $\mathbf{a}^*$ direction \cite{StoweJSolidStateChem1998,YokoyaPRB2005}. The CDW order is related to the nesting of the quasi-1D sheets that involve the 5$p$ bands of Te(2) and Te(3) [Fig.~\ref{Fig1}(a)]. The electronic gap associated with the CDW order opens near the D-point of the BZ \cite{YokoyaPRB2005,HoeschPRL2019}, where electron-phonon coupling is strongest \cite{HuPRB2015}.

\begin{figure*}
\includegraphics[width=4.5in]{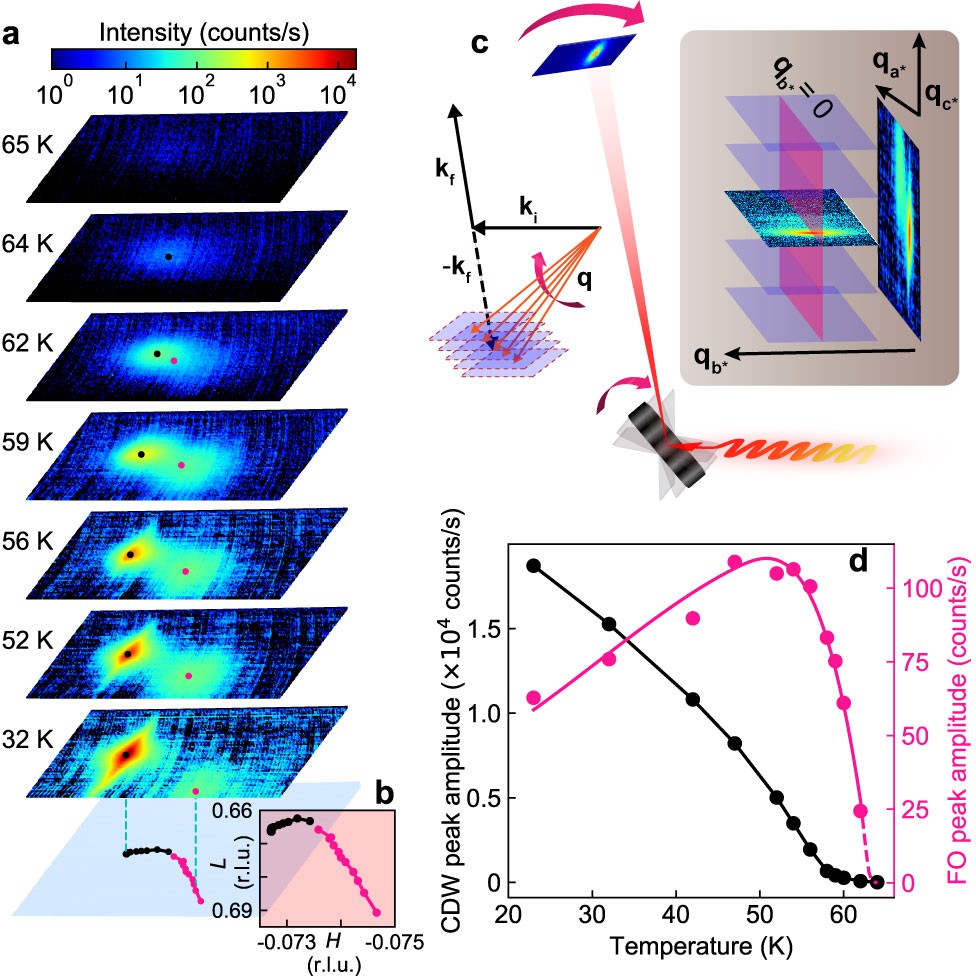}
\caption{\label{Fig2}
(a) Diffraction signals in the $[H,0,L]$ plane at various temperatures. (b) Momentum-space trajectory of the two peak centers indicated at the bottom of (a). (c) Schematics of momentum scan (see text) in real and reciprocal space. The inset displays how the $(H, K, L)$ volume data were reconstructed and the $(H, 0, L)$ slice extracted. (d) Temperature dependence of the amplitudes of the sharp (CDW) and broad (FO) peaks, determined from two-dimensional fits to the data (Fig.~S3 in \cite{SM}). Lines are guide to the eye.}
\end{figure*}

\textbf{Coexistence of CDW and FO signals.} In Fig.~\ref{Fig2} we present the temperature evolution of the CDW signal's reciprocal space fingerprints. Data at each temperature were acquired by performing a reciprocal space scan along the $\mathbf{c}^*$ direction, which involved rocking the sample while simultaneously repositioning the FCCD camera in coupled fine steps, followed by reconstruction of the FCCD images into $(H, K, L)$ volume data [Fig.~\ref{Fig2}(c)]. Thermal expansion of the lattice parameters has been accounted for based on measurements of fundamental Bragg peaks in the same temperature range. As the mirror symmetry with respect to the $ac$ plane remains intact through the CDW transition, in Fig.~\ref{Fig2}(a) we visualize the $T$ evolution of a thin $(H, L)$ slice taken near $K=0$. We make a few observations here:

(1) A clear signal persists up to at least 64 K $>T_\mathrm{CDW}$, whereby the ordering temperature $T_\mathrm{CDW}$ is determined from resistivity to be 63 K (Fig.~S1 in \cite{SM}), or from the diffraction intensity to be between 54 and 59 K [Fig.~\ref{Fig2}(d), and Figs.~S3 and S4 in \cite{SM}]. At even higher temperatures, the diffraction pattern becomes smeared and weak to the point of being buried under the fluorescence background. Nonetheless, we propose that the total signal, integrated over a reasonable finite range near $\mathbf{q}_\mathrm{CDW}$, remains significant until a much higher temperature is reached. This temperature was estimated to be $\approx 140$ K based on initial opening of electronic gaps \cite{HuPRB2015,YokoyaPRB2005}.

(2) Almost immediately below $T_\mathrm{CDW}$, two coexisting diffraction peaks materialize, subsequently evolving to develop very different momentum widths at low $T$ [Fig.~\ref{Fig2}(a), note the logarithmic color scale]. Both peaks are centered at $K=0$ independent of $T$, but they move in almost opposite directions in the $(H, L)$ plane upon further cooling [Fig.~\ref{Fig2}(b)].

(3) The two peaks exhibit distinct $T$ evolution of their amplitudes [Fig.~\ref{Fig2}(d)]. The sharp peak increases upon cooling in a monotonic order-parameter-like fashion and dominates the total signal below 59 K. In contrast, the broad peak reaches its maximum at about 50 K and then becomes gradually suppressed at lower $T$.

\begin{figure*}
\includegraphics[width=6.6in]{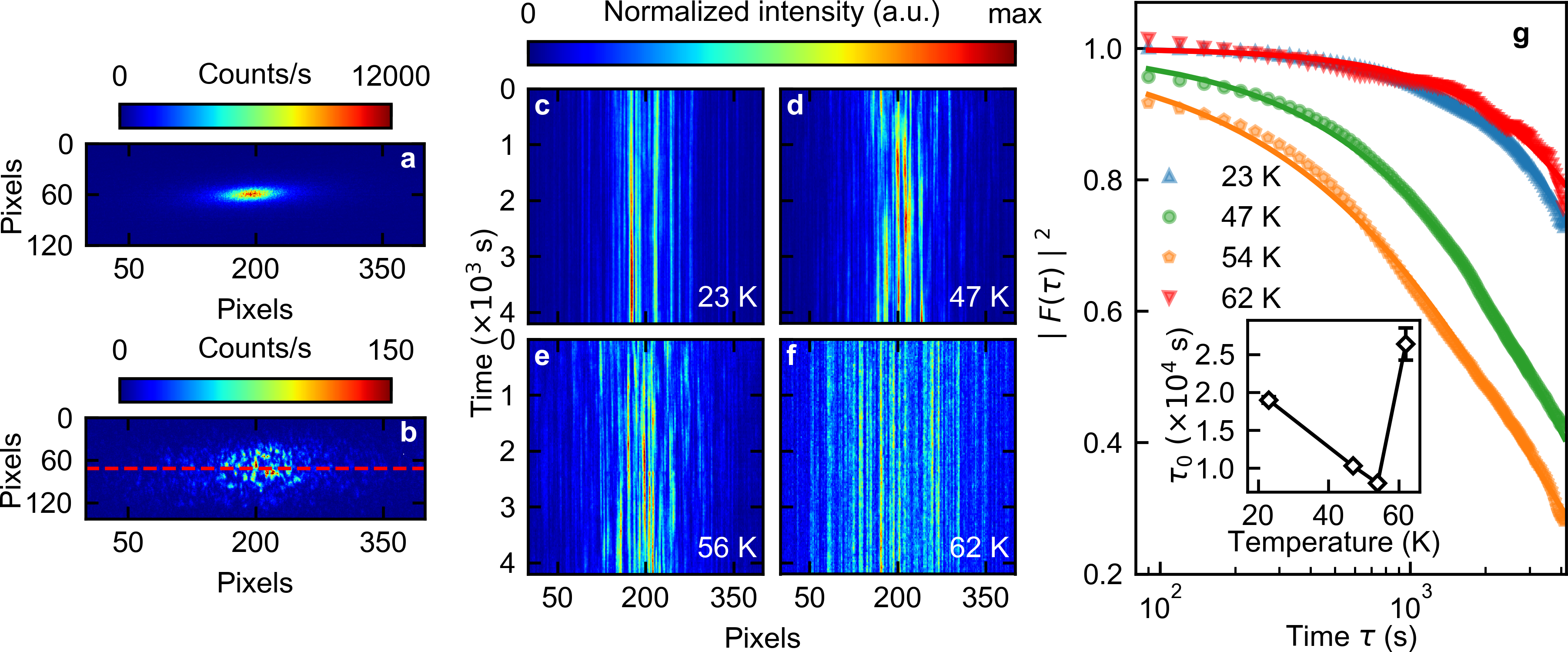}
\caption{\label{Fig3}
(a-b) Images of CDW diffraction signal taken at 45 K without (a) and with (b) the 10 $\mu$m pinhole (see text). Due to the much lower count rate in (b), the image is obtained from an average of 3 frames, each with 10 seconds of exposure time. (c-f) ``Waterfall'' plots of the time series of diffraction intensities extracted from a horizontal strip taken near the dashed line in (b). Global intensity variations over time (Fig.~S6 in \cite{SM}) have been compensated by normalizing the mean intensity of each time frame. (g) Autocorrelation of speckle patterns at different temperatures. The inset reports fitted values of the coherence time $\tau_0$ (see text) which characterizes the rate of domain motion. The autocorrelation at 62 K may exhibit some initial rapid decay (Fig.~S7 in \cite{SM}) that is not accounted for by the fit.
}
\end{figure*}

The presence of two distinct signals near $\mathbf{q}_\mathrm{CDW}$ is a surprising yet robust result. The phenomenon has been observed in two different samples that produced sufficiently sharp peaks, and the intensity of the broad peak seems to vary among samples, which may indicate a connection to sample-specific inhomogeneities, such as the nature and density of defects and/or impurities. In a conventional diffraction experiment (\textit{e.g.}, using a point detector), only the sharp peak would be noticed since it dominates the signal. Here we attribute the weak, broad peak to FO, or standing waves created by self-interfering electrons scattered off a localized potential likely arising from quenched disorder. This interpretation is consistent with the fact that the broad peak initially emerges together with the sharp peak, as they are linked to the same Fermi surface nesting instability above $T_\mathrm{CDW}$. The departure of the two wave vectors from each other upon the development of the CDW order and the associated electronic gap \cite{YokoyaPRB2005,HoeschPRL2019} can be qualitatively understood by considering various aspects: (i) the curvature of the Fermi surface; (ii) the gradual opening of the electronic gap on only part of the Fermi surface [Fig.~\ref{Fig1}(d)]; (iii) the $T$ dependence of the gap size; and (iv) the unequal contribution of different bands to the CDW and FO modulations. Moreover, our interpretation naturally explains why the broad peak retains its shape even deep in the CDW-ordered state, as well as why its highest intensity is reached at an intermediate $T \approx 50$ K [Fig.~\ref{Fig2}(d)]: the amplitude of FO depends on the coherence length of individual Bloch waves scattering off localized impurities, which improves with cooling, as well as on the density of electrons left available by the CDW gap, which instead decreases at low temperatures. The abrupt increase of the FO peak with cooling below 63 K [Fig.~\ref{Fig2}(d)] points to a positive feedback mechanism between the coupled lattice and electronic degrees of freedom, possibly enhancing the disorder scattering potential by further distorting the lattice around impurities and defects. The strong momentum dependence of such feedback mechanism \cite{HuPRB2015} may explain why the FO signals materialize in the form of a reciprocal space peak, rather than a contour \cite{KangNatPhys2019}. As we show in Figs.~S4 and S5 in \cite{SM}, an order-parameter-like behavior and long correlation lengths of the CDW are only realized below 56$\pm3$ K, so the resistivity anomaly at 63 K (Fig.~S1 in \cite{SM}) is probably caused by the abruptly enhanced disorder scattering, rather than by the CDW ordering itself.

Related to our above observation, non-resonant X-ray diffraction signals associated with FO have previously been reported in the quasi-1D CDW material K$_{0.3}$(Mo$_{0.972}$V$_{0.028}$)O$_3$ with controlled disorder introduced through vanadium impurities \cite{RouzierePRB2000}. However, the high impurity concentration did not allow the previous authors to separate the CDW and FO signals, nor to address their relation on the verge of CDW formation. In a more recent study of heavy-fermion compounds \cite{GyenisSciAdv2016}, Gyenis \textit{et al.} demonstrated that resonant X-ray diffraction indeed possesses the sensitivity to detect FO, the signal of which exhibited resonant enhancement and increased towards low $T$ similar to our results. However, the absence of CDW order in those materials precluded a differentiation of CDW and FO diffraction signals, a methodology that is most needed in our context \cite{TorreNatPhys2016,TorrePRB2016,TorreNewJPhys2015}. To this end, our direct observation of coexisting CDW and FO signals, along with their distinct temperature evolution, is new and revealing. Hereafter, we present another method to distinguish the two effects.

\begin{figure*}
\includegraphics[width=4in]{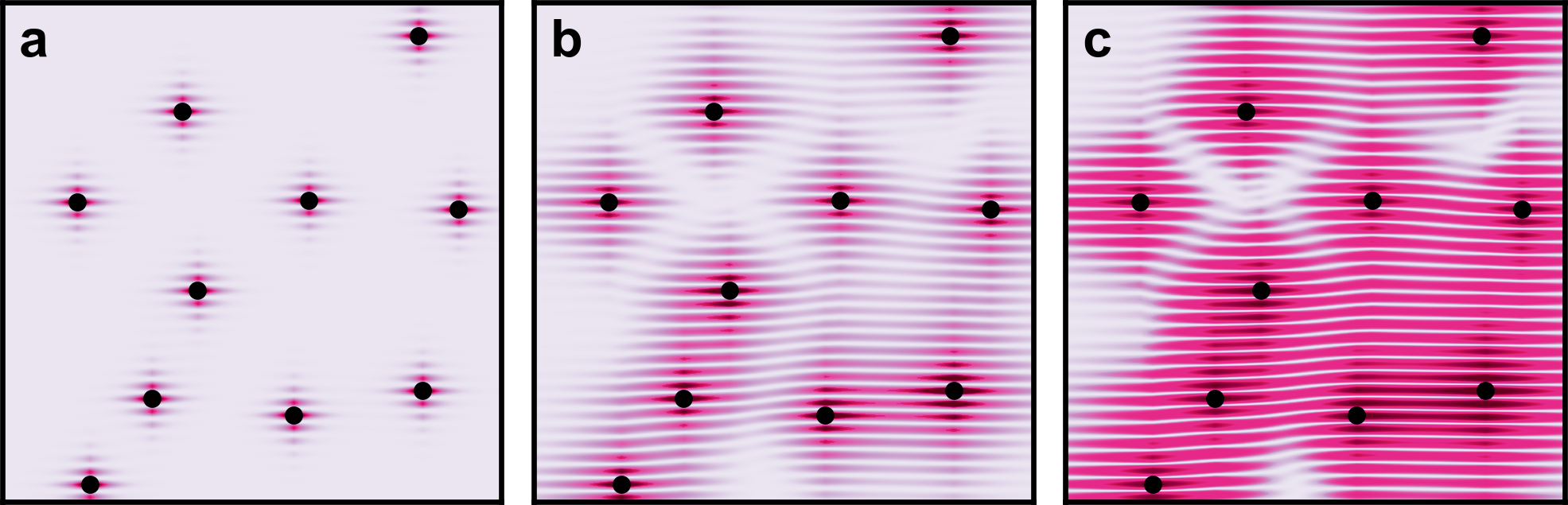}
\caption{\label{Fig4}
Schematics of the charge modulations in the presence of disorder (black dots) in different temperature regimes: (a) $T > T_\mathrm{CDW}$, (b) $T \lesssim T_\mathrm{CDW}$, and (c) $T \ll T_\mathrm{CDW}$.
}
\end{figure*}

\textbf{Mesoscopic dynamics.} Figure~\ref{Fig3}(a-b) displays a comparison of FCCD images taken before and after insertion of a 10 $\mu$m pinhole into the X-ray beam's path about 5 mm upstream from the sample. The pinhole significantly improved the beam coherence by reducing the beam size (and hence the illuminated sample volume). As a result, we observed speckle patterns [Fig.\ref{Fig3}(b)] due to the interference of the X-rays diffracted from different CDW domains. The beamline's high photon flux allowed us to record a statistically significant speckle pattern within less than a minute of photon accumulation, and its high stability allowed us to monitor the pattern over a time span up to hours. Dynamical charge domains manifest themselves as time-varying speckle patterns. In Fig.~\ref{Fig3}(c-f), we present ``waterfall'' plots of the speckle pattern time series recorded at several temperatures, which are constructed by vertically stacking narrow horizontal sections [11-pixel wide, see dashed line in Fig.~\ref{Fig3}(b)] of FCCD data acquired at different times. The continuous and straight vertical streaks in Fig.~\ref{Fig3}(c) and (f) indicate that the speckle patterns are very stable at 23 K and 62 K, whereas the broken streaks in Fig.~\ref{Fig3}(d) and (e) suggest presence of mobile CDW domains at these intermediate temperatures. The same conclusion can be captured more quantitatively by analyzing the intensity autocorrelation function \cite{ChenPRL2016} of the speckle patterns:
\begin{equation}\label{Eq1}
g_2(\tau)\equiv\frac{\left \langle I(p,t) I(p,t+\tau) \right \rangle} {\left\langle I(p,t)\right\rangle ^2}\equiv1+\beta|F(\tau)|^2,
\end{equation}
where $I(p,t)$ is the intensity obtained at time $t$ and pixel $p$, $\tau$ is the time difference, and $\beta$ is a measure of the contrast of the speckle patterns. The average $\left\langle\cdots\right\rangle$ is taken over $t$ and all pixels in a small region near $\mathbf{q}_\mathrm{CDW}$. $F(\tau)$ is the intermediate scattering function, for which we assume an exponential form:
\begin{equation}\label{Eq2}
|F(\tau)|=e^{-(\tau/\tau_0 )^\alpha},
\end{equation}
where $\tau_0$ represents the characteristic time required for the domain distribution to change significantly, and $\alpha$ is the ``stretching exponent'' \cite{ShpyrkoNature2007}, determined to be around unity in our experiment (Fig.~S7 in \cite{SM}). It can be seen that the domains are less static at 47 and 56 K than at 23 K, which can be explained by thermal activation of domain walls. However, upon further heating to 62 K, which is close to or even slightly above the melting point of the CDW order, the domains become static again.

The presence of anomalously static domains when the CDW order is about to melt cannot be attributed to thermal effects. A plausible explanation is that these ``domains'' are essentially disconnected FO, similar in essence to the anomalous static ``central peak'' observed above structural phase transition temperature in SrTiO$_3$ \cite{HoltPRL2007}. As we illustrate in Fig.~\ref{Fig4}(a), such a situation can be realized in the dilute-disorder limit. In this scenario, there is an appreciable temperature range above $T_\mathrm{CDW}$ where FO are sufficiently strong to produce a resonant X-ray diffraction signal, but the FO patches formed around different sites remain isolated from each other.  Upon cooling below $T_\mathrm{CDW}$, the correlation length rapidly grows, and it is at this point that regular domain walls form [Fig.~\ref{Fig4}(b)]. As long as the domain walls are not too rigid to be thermally perturbed, domain redistribution is expected to occur over time, consistent with our observations at 47 and 56 K in Fig.~\ref{Fig3}. At very low temperatures, the CDW fabric becomes stiff, and its domain walls are robust against thermal perturbations [Fig.~\ref{Fig4}(c)].

\section{Discussion}

The scenario invoked in Fig.~\ref{Fig4}(a) to explain our coherent-scattering data obtained at 62 K [Fig.~\ref{Fig3}(f,g)] is consistent with the fact that this temperature marks a rapid growth of the FO diffraction signal in Fig.~\ref{Fig2}(d) that is not accompanied by a dominating CDW signal. In fact, even for the CDW peak in Fig.~\ref{Fig2}, a detailed analysis suggests that very long correlation lengths are reached only below $T\approx56$ K along with a saturation in the ordering vector (Figs.~S3 and S5 in \cite{SM}), consistent with our observation of mobile domain walls at 47 and 56 K. Based on these results, we therefore believe that the dilute disorder sites in our ZrTe$_3$ samples act as pinning centers \cite{GrunerRevModPhys1988,NovelloPRB2015,LebollochPRB2005} seeding the CDW formation. They first generate disconnected FO nanopatches, which later merge to pave the way for long-range CDW order at low $T$, and also lead to unstable domain walls at intermediate $T$. The disconnected FO patches are fully static because the disorder sites have strong scattering (and pinning) potential, which can be further enhanced by the aforementioned positive feedback below 63 K. Our interpretation does not require a direct pinning of domain walls by disorder, since in the dilute-disorder limit, all domain walls are expected to end up in pristine regions of the crystal [Fig.~\ref{Fig4}(c)]. Importantly, since the CDW order does not lead to a full gap, FOs arising from ungapped portions of the Fermi surface remain present as an incipient instability (Fig.~\ref{Fig2}) below $T_\mathrm{CDW}$.

Our results suggest that dilute quenched disorder plays a seeding role for CDW formation. Moreover, their presence helps define a genuine CDW phase transition in experiments, since they give rise to distinct observables [Figs.~\ref{Fig2} and \ref{Fig3}] for the CDW and FO phenomena. In the light of our work, such a clear-cut separation appears to be missing in a recent study of the La$_{1.875}$Ba$_{0.125}$CuO$_4$ cuprate superconductor \cite{ChenPRL2016}, where the observed diffraction signal is not as sharp as in our case, and the speckle patterns are found to be static at all temperatures. The counterpart of our finding, namely, the role of disorder in charge-order formation in the strong-coupling limit, to which the cuprates belong, deserves further experimental scrutiny.



\section{Methods}
\textbf{Calculation of the Fermi surface.} Each cut of the Fermi surface on $\mathbf{a^{*}b^{*}}$ plane was calculated  using Quantum Espresso package~\cite{Giannozzi2009}  based on density-functional theory, within the generalized-gradient approximation parameterized by Perdew, Burke and Ernzerhof \cite{Hedin1971,Perdew1996}. Norm-conserving pseudopotentials, generated by the method of Goedecker, Hartwigsen, Hutter, and Teter \cite{Hartwigsen1998}, were used to model the interactions between valence electrons and ionic cores of both Zr and Te atoms. The Kohn-Sham valence states were expanded in the plane wave basis set with a kinetic energy truncation at 150 Ry.  The equilibrium crystal structure was determined by a conjugated-gradient relaxation, until the Hellmann-Feynman force on each atom was less than $0.8\times10^{-4}$ eV/$\mathrm{\AA}$ and zero-stress tensor was obtained.  A 12$\times$18$\times$8 $\mathbf{k}$-grid centered at the $\Gamma$ point was chosen in the self-consistent calculation, following by a non-self-consistent calculations on a $\mathbf{k}$-grid of 54$\times$78$\times$1  to obtain the  Fermi surfaces. A Gaussian-type broadening of 0.0055 Ry was adapted.

\textbf{Scattering experiment.} Single crystals of ZrTe$_3$ were synthesized by a vapor transport method \cite{HuPRB2015}. Resonant soft X-ray scattering experiments were performed at the NSLS-II facility (Brookhaven National Laboratory) on beamline 23-ID-1, which provides coherent X-rays with a high flux of about $10^{13}$ photons/sec and excellent mechanical stability. A fast charge-coupled-device (FCCD) camera with a maximal readout rate of $100$ Hz and $30\times30$ $\mu$m$^2$ pixel size was placed $34$ cm away from the sample, which was mounted with the $[H, 0, L]$ reciprocal plane lying in the vertical scattering plane. The measured CDW signals were located at about $(-0.07, 0, 0.67)$ in reciprocal lattice units (r.l.u.), a satellite reflection near the $(0, 0, 1)$ Bragg peak. The horizontally polarized incident beam was tuned to a photon energy of 630 eV, in order to maximize the CDW diffraction signal (Fig.~S2 in \cite{SM}).

\textbf{Fitting of the diffraction signals.} In order to extract the temperature-dependent parameters that describe the CDW and the FO signals in Fig.~2, we use two two-dimensional Gaussian profiles with anisotropic widths. The results are displayed in Figs.~2, S3 and S4. A detailed study of the line-shape characteristics of the data, as has been done for instance in \cite{RavyPRB2006}, is beyond the scope of our present study.

\begin{acknowledgments}

We wish to thank B. Fine, B. Keimer, S. A. Kivelson, M. Le Tacon, M. Minola, Y. Y. Peng, N.-L. Wang, Yan Zhang, and Yi Zhang for discussions. The work at PKU is supported by the NSF of China under Grant Nos. 11874069 and 11888101, and by the NBRP of China under Grant Nos. 2018YFA0305602 and 2015CB921302. The work at MIT is supported by the National Science Foundation under Grant No. 1751739. The work at Jinan University is supported by the NSF of China under grant No. 11804118. This research used beamline 23-ID-1 of the National Synchrotron Light Source II, a U.S. Department of Energy (DOE) Office of Science User Facility operated for the DOE Office of Science by Brookhaven National Laboratory under Contract No. DE-SC0012704. Part of the calculation was supported by High Performance Computing Cluster in Jinan University.

\end{acknowledgments}

\bibliography{Reference}

\end{document}